# Radar, without tears

François Le Chevalier[1]

May 2017

*Originally invented to facilitate maritime navigation and avoid icebergs, at the beginning of the last century, radar took off during the Second World War, especially for aerial surveillance and guidance from the ground or from aircraft. Since then, it has continued to expand its field of employment.*

## 1   What is it, what is it for?

Radar is used in the military field for surveillance, guidance and combat, and in the civilian area for navigation, obstacle avoidance, security, meteorology, and remote sensing.

For some missions, it is irreplaceable: detecting flying objects or satellites, several hundred kilometers away, monitoring vehicles and convoys more than 100 kilometers round, making a complete picture of a planet through its atmosphere, measuring precipitations, marine currents and winds at long ranges, detecting buried objects or observing through walls – as many functions that no other sensor can provide, and whose utility is not questionable. Without radar, an aerial or terrestrial attack could remain hidden until the last minute, and the rain would come without warning to Roland–Garros and Wimbledon ...

Curiously, to do this, the radar uses a physical phenomenon that nature hardly uses: radio waves, which are electromagnetic waves using frequencies between 1 MHz and 100 GHz – most radars operating with wavelength ($\lambda = c / f$, $\lambda$ being the wavelength, c the speed of light, $3.10^8$ m/s, and f the frequency) between 300 m and 3 mm (actually, for most of them, between 1m and 1cm): wavelengths on a human scale ... Indeed, while the optical domain of electromagnetic waves, or even the ultraviolet and infrared domains, are widely used by animals, while the acoustic waves are also widely used by marine or terrestrial animals, there is practically no example of animal use of this vast part, known as "radioelectric", of the electromagnetic spectrum (some fish, Eigenmannia, are an exception to this rule, but they use frequencies on the order of 300 Hz, and therefore very much lower than the "radar" frequencies). Generators, receivers, antennas, etc. therefore have no equivalent in nature, and had to be invented « from scratch ».

The main interest of this "radioelectric" frequency band is that they easily propagate in the atmosphere, even in the presence of rain or fog or smoke: Figure 1 shows that the attenuation is less than one dB/km (Remind : 1dB is a factor 1.25, 3dB is a factor 2, 10dB is a factor 10) over almost all this frequency band – the lowest frequencies being considerably less attenuated. Moreover, since the corresponding wavelengths are of the

---

[1] Many thanks to my colleagues who reviewed and improved an earlier version of this essay : Y. Blanchard, E. Chamouard, C. Enderli, F. Gosselin, N. Petrov, L. Savy

centimeter type, the antennas remain of reasonable dimensions, as will be seen hereinafter.

Overall, this same part of the electromagnetic spectrum, whose propagation and reflection properties were highlighted by Hertz at the end of the 19th century, are used by all modern means of communication (radio, communications, localization). Nowadays, all of these radioelectric means are used, often jointly, to carry out the main tasks of surveillance, navigation and guidance; For example, for air traffic control, satellite positioning systems (using radio waves at 1 GHz) are complemented by surveillance radars (often around 3 GHz), for airport security, and weather radars (at 3, 5, or 10 GHz) for weather monitoring.

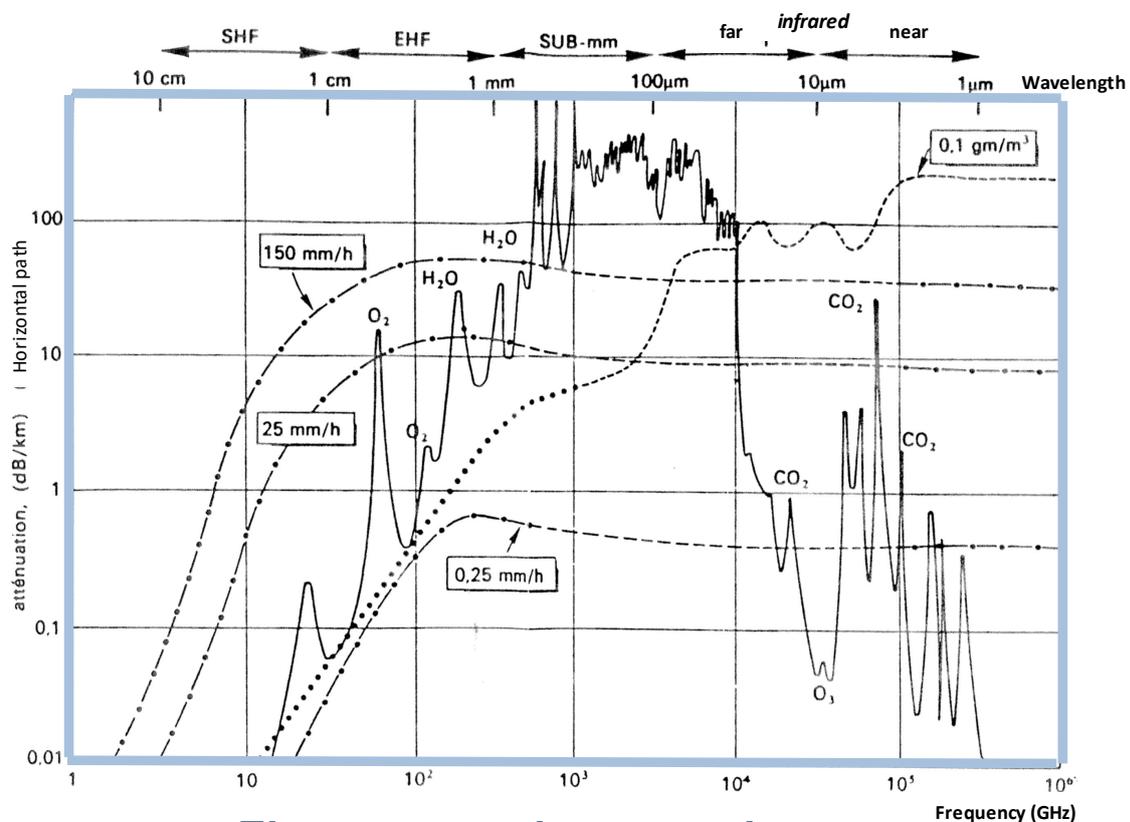

– **Figure 1** : *Electromagnetic propagation in the atmosphere* –

## 2  How does it work?

### 2.1  Principle

The principle of radar is that of the echo: a wave is created and sent, propagates in the vacuum or in the atmosphere, then is reflected by the object – the "target" in the martial vocabulary that radar has inherited from its warlike origins – and captured in return. This principle, well known in acoustics as sonar (and used by bats, dolphins, and … mountaineers), is also valid in electromagnetism, and in particular for these "radioelectric" waves, since most of the obstacles, natural or artificial, send back a radioelectric echo – it is indeed difficult to reduce this echo, as the modern "stealth" technology demonstrates.

It is thus an "active" system, implementing a transmitter and a receiver, according to the principle diagram represented in Figure 2: an electric oscillation, of roughly sinusoidal shape, is created (waveform generator), amplified and shaped by a transmitter: until then, one remains in the domain of electrical (or electronic) circuits. This signal is then transmitted by the antenna to the external environment in the form of an electromagnetic wave which propagates towards the target and then returns (partly, because it is strongly attenuated by propagation and by the reflection on the target) and is detected by the receiving antenna (generally common with the transmitting antenna), and transmitted again to the receiver as an electrical signal. This receiver compares the very small received signal with the transmitted signal (to check that it is indeed an echo, and to extract it from the surrounding noise), and transmits this information to a display console or, more generally, to the system in charge of exploitation of the information.

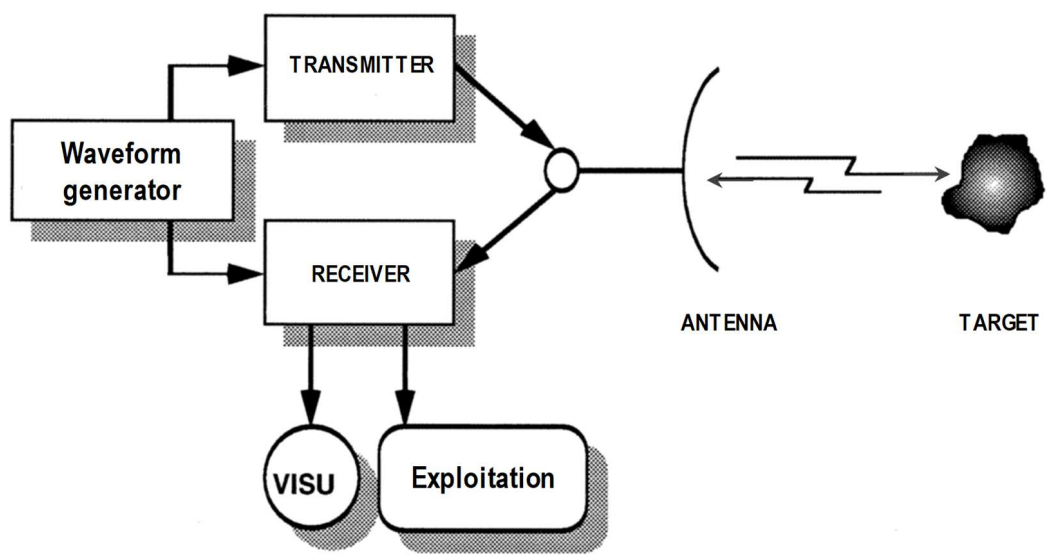

− **Figure 2** : *Schematic diagram of a radar* −

The presence of this echo provides, besides the index of presence of a target ("detection"), its distance, by measuring the delay between transmission and reception (extremely short delay, given the high speed of the light: $c = 3 \cdot 10^8$ m/s: a delay of 1µs for the round trip on a target at 150m), and its direction, by the knowledge of the angular sector illuminated by the antenna. RAdio Detection And Ranging, RADAR – one could then add "and angular localization", but also "and velocity measurement", because the radar measures the speeds of the targets thanks to the Doppler effect: indeed, this measurement of speed is essential to separate the "fixed" echoes (on buildings, trees, mountains, etc.) from "mobile" echoes, which are generally echoes of interest to the user.

## 2.2 Doppler effect

The Doppler effect, which indeed plays a crucial role in radar, can be understood simply by considering a sinusoidal wave of frequency f (ie f oscillations per second) emitted by a fixed transmitter and propagating at a speed c – with a wavelength $\lambda = c/f$ – and received by a sensor traveling at a velocity v relatively to the transmitter: due to its regular movement towards the source, the sensor will see a duration of each oscillation (period) slightly shorter; More precisely, in 1s, it will count a number of oscillations no longer equal to f, but to $f+v/\lambda$ : indeed, from the point of view of the target, the apparent speed of the wave is c+v : the apparent covered distance by the wave during 1s is c*1s + v*1s hence a count of oscillations, each of length lambda, given by $c/\lambda + v/\lambda = f + v/\lambda$. In other words, if the sensor has approached k wavelengths in 1s, it will count a number of oscillations equal to f + k.

The Doppler effect is thus an apparent shift of the emitted frequency equal to $v/\lambda$ (ie equal to the radial velocity – the velocity component parallel to the direction of propagation – expressed in wavelengths per second), positive if the sensor moves to the transmitter, and negative if not. It does not depend on the nature of the wave which is transmitted, and is also observed for sound waves, hence the well–known effect on sirens of ambulances ... This frequency shift occurs on the path from the antenna to the target, but also on the return path, resulting in a total shift of $2v/\lambda$ of the radar frequency. In order of magnitude, this corresponds to an offset of 200 Hz for a target traveling at a radial speed of 10 m/s (36 km/h), as seen by a radar operating at 10 cm wavelength (at a carrier frequency of 3 GHz): this shift is therefore significant, but very delicate to measure, since the relative accuracy required is $200/(3.10^9)$, or $6.10^{-8}$: to achieve such an accuracy, the generated sinusoid must be extremely pure, which leads radarists to design systems built on a single clock, of high spectral purity, from which are generated all the necessary signals: pulse cutting, signal sampling, local oscillators, etc.

With these orders of magnitude, it can be seen that not only, as mentioned above, the transmitters, receivers, antennas, etc. have no equivalents in nature, but also that the quality of the requirements required to perform a correct detection is very high.

Thus, the radar operates on the principle of echo using radio waves, these echoes being affected by a Doppler effect which makes it possible to measure their velocities. Before constructing a generic radar, we still have to say a word about the antenna, which is an essential component of the radar, – the most visible on the outside – , and which is a key factor for the "power budget", ie to determine the required power and the corresponding maximum ranges.

## 2.3 The antenna

As mentioned above, the antenna transforms the electrical signal leaving the transmitter into a wave propagating in the surrounding environment, in a certain angular sector – and ensures the inverse transformation on receive. It thus fulfills two functions: the adaptation between the waveguide or the coaxial cable carrying the electrical signal and the external environment, and the focusing of the emitted wave in a certain (usually narrow) angular sector.

For "centimeter" waves – typical of radar – , a quasi–optical approximation can be used to analyze the propagation of waves: this approximation makes it possible to understand simply the operation of the parabolic antennas of traditional radars (Figures 3 and 4): The waveguide disposed after the transmitter is terminated by a horn which provides the adaptation with the external medium. This horn is placed at the focus of a parabola which makes it possible to generate a plane wave propagating in the direction of the xx' axis of the parabola.

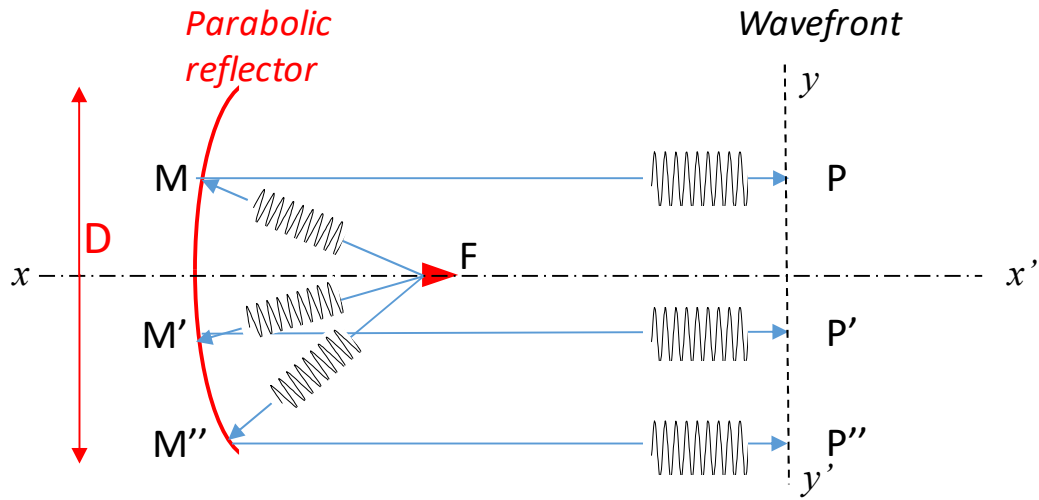

– **Figure 3** : *Parabolic reflector antenna* –

Indeed, by virtue of the basic geometrical property of the parabola, the paths FMP, FM'P', FM''P'', etc. are of the same length, so that the signal emitted from the focus F simultaneously arrives at P, P', and P'': yy' defines the trace of the wave front (perpendicular to the propagation axis xx' ) in the figure, and this reflector antenna emits a plane wave in the direction of its xx' axis.

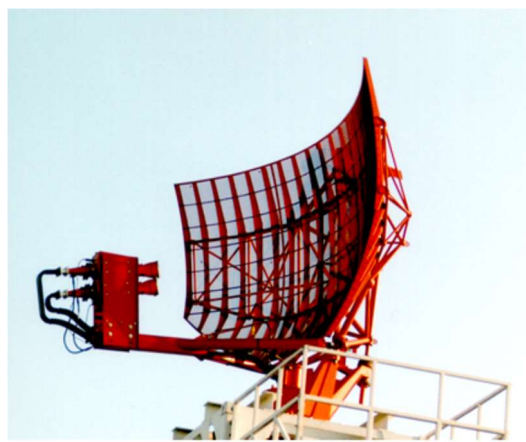

– **Figure 4** : *Rotating antenna with reflector, to monitor air traffic in the terminal area (range: 100 km) of an airport; We note the presence of 2 horns in the vicinity of the focus, to create 2 beams slightly offset in elevation* –

In fact, this wave is emitted in a certain angular sector (cone) around the direction x x'. It can be shown that the opening of this angular sector, expressed in radians, is equal to λ / D, λ being the wavelength used and D the dimension of the antenna: in other words, the antenna ability to focus the emitted energy is directly proportional to its dimension, expressed in wavelengths. Similarly, for a two–dimensional antenna, one obtains an ability to focus in azimuth proportional to the width of the antenna, and in elevation proportional to its height:

$$\theta_{azimuth} = \frac{\lambda}{D_{horizontal}} \quad , \quad and \quad \theta_{elevation} = \frac{\lambda}{D_{vertical}} \; ;$$

Typically, for a wavelength of 10 cm and an antenna dimension of 2 m, an angular selectivity of the order of 1/20 radian is obtained, ie approximately 3°: this allows an approximate angular localization, but with limited resolution (1/20 radian is equivalent to 5 km at 100 km range!). We see that it is interesting to choose small wavelengths to improve this angular resolution – but then, it is the propagation losses (Figure 1) that limit the maximum range of the radar ... This compromise between angular resolution and propagation losses explains that short–range radars generally use small wavelengths, between 3 mm and 3 cm, especially for applications where the available volume is severely constrained, on board aircraft or vehicle in particular. Conversely, for long range radars, larger wavelengths, typically between 10 cm and 1 m, are generally chosen and large antennas are used.

Nowadays, this rotating parabola for successively exploring the different directions is increasingly replaced by an "active antenna" network, according to Figure 5: rather than a wave emitted by a single source and then reflected by a parabola, one uses a quantity of elementary sources (typically one thousand) transmitting simultaneously the same signal: this produces in the same way a plane wavefront of trace P P'P''  in the figure, perpendicular to the axis x x', with the same angular selectivity (λ/*D*) as in the case of a parabola.

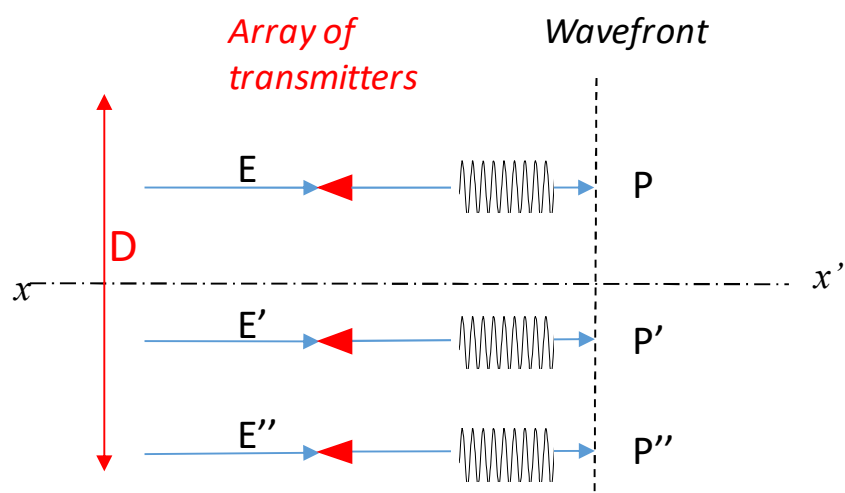

– **Figure 5** : *Array antenna, boresight looking*  –

The advantage of this solution is that if there are suitable programmable delays before the transmitters, τ, τ', τ'', etc., it is possible to orient the wave front in any direction, according to Figure 6: it thus becomes possible to carry out a very agile "electronic scanning" without any mechanical movement. Another advantage is that the single high-power transmitter placed at the focal point of the parabola has been replaced by a series of low-power elementary transmitters which can be realized in solid state technology: this system is more tolerant of elementary failures, and now feasible at an acceptable cost thanks to the series effects (1 antenna = 1000 transmitters / receivers, typically).

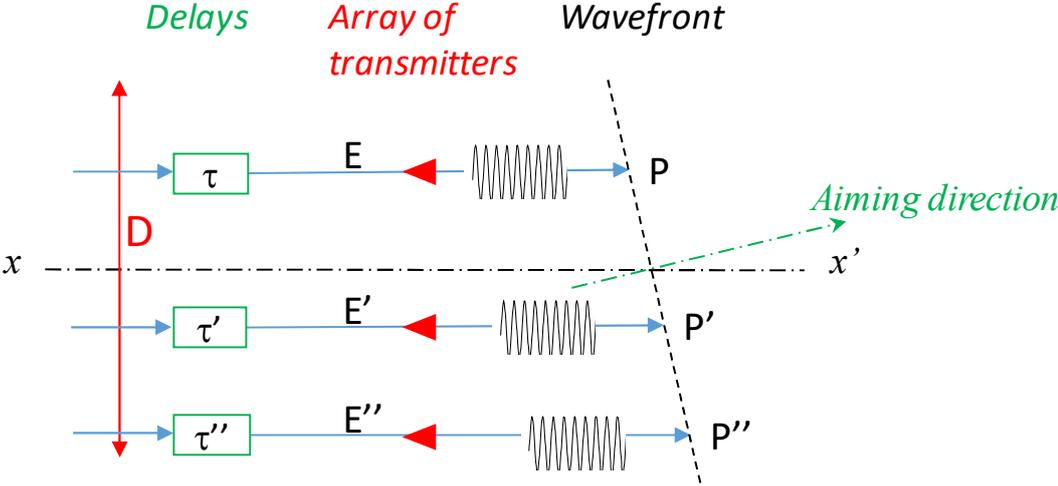

– **Figure 6** : *Array antenna, squint looking* –

Two examples of radars using this so-called active-transmitting-receiving antenna technique are shown in Figure 7: a rotating antenna, combining mechanical scanning and electronic scanning for sectorial observation (COBRA radar for counterbattery), and a fixed antenna, for airborne front radar (RBE2 radar of the Rafale aircraft).

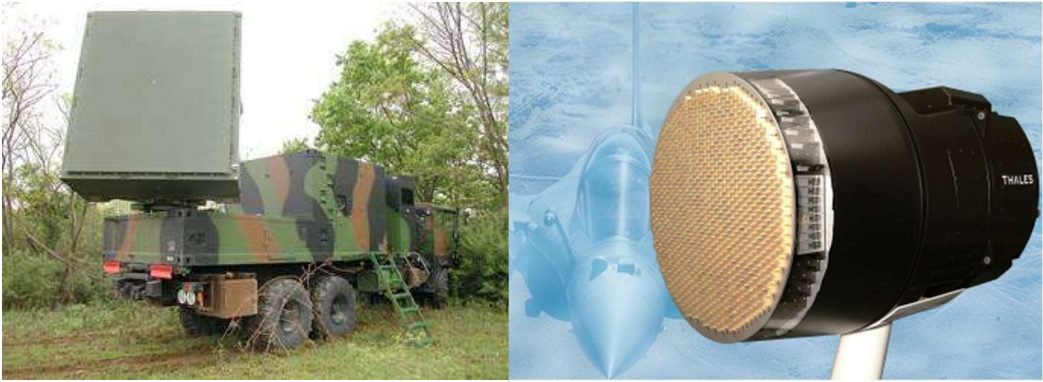

– **Figure 7** : *Modern radar active antennas (Courtesy of Thales): on the left, an active electronic scanning antenna (COBRA radar), and on the right an airborne fixed electronic scanning antenna (RBE2 radar of the Rafale)* –

## 2.4  The waveforms

Since the received echo is very low compared to the transmission, it is necessary to shut the transmission during reception (not easy to listen while screaming...). Moreover, to measure the distance with precision, it is obviously preferable to use a pulse as short as possible. Most radars therefore transmit short pulses of the order of one microsecond (if the duration of the pulse is 1 microsecond, then its "length" is 300 m, that is to say 150 m taking into account the two–way path of the wave). For a long–range radar, for example 150 km, the echo returns after 1 ms: it is therefore possible to transmit these pulses with a repetition period of at least 1 ms. Such a radar then typically transmits pulses with duration 1 µs, on a carrier around 3 GHz (leading to 3000 oscillations per pulse : it looks indeed like a portion of a continuous wave ! See Figure 9 below)), repeated every ms.

In the real world, the received echo is in fact the superposition of many distinct echoes contained within the 150m range resolution cell : a real target produces many elementary echoes (coming from reflexions from different parts, eg wings, fuselage, cockpit, etc.) summing up to an equivalent echo, itself being mixed up with ground echoes from the same resolution cell as the antenna beam may illuminate the ground as well as the target : the Doppler effect can help discriminating the useful echoes : actually, one carrier is emitted, but multiple (slightly different) frequencies are received from the moving targets – and then separated on receive, by spectral analysis, just as different notes are separated in audio.

The measurement of the velocity by the Doppler effect is not possible with sufficient accuracy (on the order of 10 to 100 Hz, typically, as described in paragraph 2.2) over the duration of a pulse: actually, over a duration T, it is possible to measure a frequency shift δf with a resolution 1/T, ie it is possible to measure one more (or one less) oscillation: $2\pi \, \delta f \, T \geq 2\pi$ (see Figure 8). The Doppler will therefore be measured by analysis of a train of 10 to 100 consecutive pulses – and therefore the radar antenna must remain pointed (mechanically or electronically) in each direction for typically 10 to 100 ms.

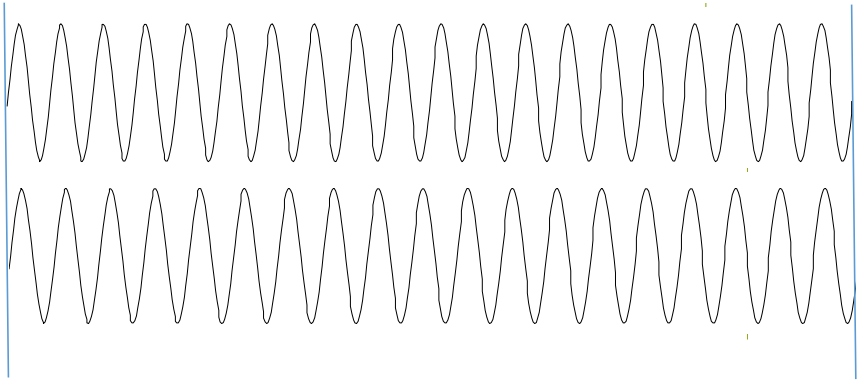

– **Figure 8** : *Two discernible sinusoids (for example counting zero crossings): 20 oscillations for the first, 19 for the second –*

It is thus seen that a fine analysis of speed requires a rather long observation time in each direction, and therefore a rate of renewal of the information not too high: this is one of the compromises to which the radar must comply (such as that mentioned above linking the angular resolution and the maximum range in rainy weather, via the choice of the wavelength).

Another compromise appears quickly with this type of pulse waveform: improving the range resolution leads to shortening the emitted pulses and thus reducing the energy sent to the target[2] – and thus the energy received from this target: it then becomes difficult to obtain a good resolution in range, necessary to separate close planes, for example, at long distance. Fortunately, there is a way out of this dilemma, by modulating the pulse (phase or frequency modulation): instead of emitting a "pure" sinusoid, a sinusoid modulated in frequency is emitted, (for example a linear modulation, for the signal known as « Chirp » – which, by the way, is a typical waveform used by bats for ultrasound echolocation), which allows to decouple the pulse duration from the range resolution, because a long but very modulated pulse makes it possible to measure the distance with a good resolution. Without going into details, however, it should be noted that on receive, the signal processing will obviously have to take account of this a priori knowledge of the shape of the transmitted pulse, in order to extract from the received signal "what looks like the transmitted signal".

## 2.5 Power budget

Consider a radar having a transmitter with power $P_t$, and seeking to detect a target at a distance $R$, with a rectangular antenna of area $A = D_{horizontal} . D_{vertical}$, hence of angular resolutions $\theta_{azimuth} = \frac{\lambda}{D_{horizontal}}$ , and $\theta_{elevation} = \frac{\lambda}{D_{vertical}}$ ;

Let us assume first that the emission antenna is isotropic, thus illuminating $4\pi$ steradians[3]. The target is characterized by its "radar cross section", that is to say that it intercepts the fraction $\frac{\sigma}{4\pi R^2}$ of the transmitted power, $\sigma$ being the effective cross–section[4] and $4\pi R^2$ the surface of the sphere radiated from the emitter at the distance $R$. This intercepted power is re–radiated from the target in all directions and the receiving antenna in turn intercepts the fraction $\frac{A}{4\pi R^2}$. The received power is thus written :

$$P_r = P_t \frac{\sigma}{4\pi R^2} \frac{A}{4\pi R^2}$$

In practice, the transmitting antenna is not isotropic (it is generally the same as the receiving antenna), it can be characterized by its gain $G$, which simply measures its ability to focus the energy, with respect to an isotropic antenna: the gain is therefore the ratio between the $4\pi$ steradians covered by an isotropic antenna and the $\theta_{azimuth} . \theta_{elevation}$ steradians covered by the directive antenna (for small angles, the steradians are simply the product of the radians in 2 orthogonal directions), or $G = \frac{4\pi A}{\lambda^2}$. We obtain :

---

[2] Since the peak power of the transmitter is generally limited, the shortening of a pulse cannot be compensated for by increasing the peak power.

[3] According to Wikipedia, the steradian (symbol: sr) is the unit of solid angle. It is used in three-dimensional geometry, and is analogous to the radian which quantifies planar angles.

A steradian can be defined as the solid angle subtended at the center of a unit sphere by a unit area on its surface. For a general sphere of radius r, any portion of its surface with area A = r² subtends one steradian.

Because the surface area A of a sphere is 4πr², the definition implies that a sphere measures 4π (≈ 12.56637) steradians. By the same argument, the maximum solid angle that can be subtended at any point is 4π sr.

[4] Typical orders of magnitudes for radar cross sections are 1 m² for a pedestrian, a few m² for a fighter aircraft, 10 m² for a vehicle, and thousands of m² for a ship.

$$P_r = P_t G \frac{\sigma}{4\pi R^2} \frac{A}{4\pi R^2} = P_e \frac{\sigma A^2}{4\pi \lambda^2 R^4}$$

This radar transmits a pulse, which can be seen as a « portion of plane wave », illustrated on Figure 9, with widths $\theta_{azimuth}$ and $\theta_{elevation}$ – typically a few degrees each – and depth $\Delta R$ (range resolution, equal to the length of the pulse – typically 150m – , when there is no specific modulation) ; this pulse is transmitted, reflected by the target, and received with the residual power $P_r$.

The power budget is therefore proportional to $1/R^4$, decreasing very rapidly with distance. Typically, for a radar cross–section of 1 m² and a surface antenna 1 m², the power received at 10 km is about 150 dB lower than the transmitted power! If one transmits kilowatts, one receives picowatts: it is indeed necessary to shut the transmission during the reception.

It is also necessary to use an amplifier of excellent quality in reception, so as not to mask the very low received signal by the noise of this receiver: the radar is a school for high quality: high spectral purity, low noise factor, suitable signal processing, etc.

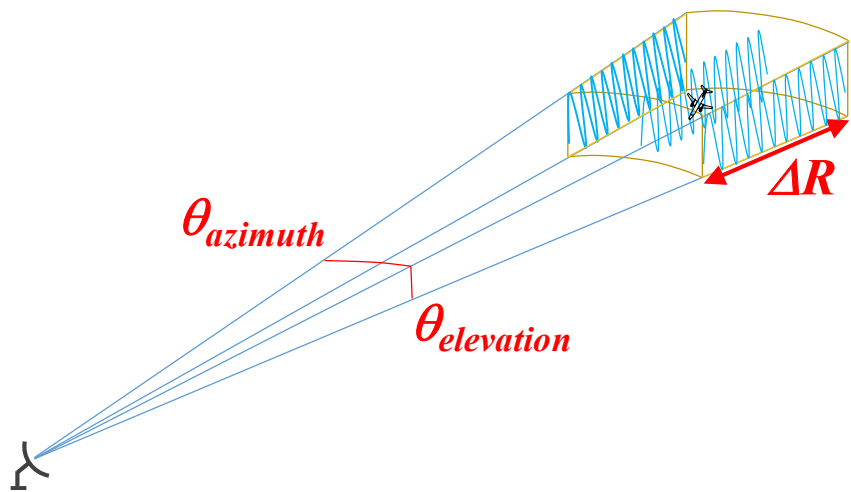

– **Figure 9** : *Baseline radar transmission* –

## 3 Future radars

The principles of radar are now well established, but the required technologies (miniaturization of transceivers and increased performance of analog–to–digital converters, in particular) are rapidly evolving: radars are now being designed with signals emitted simultaneously in different directions (Figure 10), in order to better resolve the compromises mentioned above.

There is also a trend towards widening the bandwidth of signals used, in order to improve the accuracy of targets analysis of and to discriminate, for example, a low–altitude drone from a vehicle, even in urban environments returning very strong fixed echoes.

Finally, all the available information processing resources (including multisensor fusion, and deep learning analysis for reliable identification of targets) are used to improve the management of these very agile sensors, and extract the best knowledge about the targets and their environment.

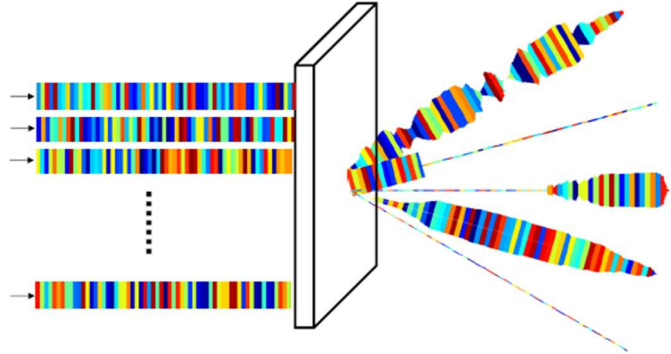

– **Figure 10 (by J.P. Guyvarch)** : *Multiple simultaneous transmissions – Multiple Input Multiple Output (MIMO), providing better angular resolution and adaptation of the radar to its environment, both for emission and for reception.*

In general, the main enemy of the surveillance radar remains fixed echoes (buildings, trees) or slowly mobile echoes (rain, foliage, sea). For this reason, the development of new sensors requires very fine simulations of environments and nuisances (including other emissions, telephony, Wifi, radar, etc.), and real-world testing to take full account of the infinite variety of real situations. These severe constraints also drive the search for high quality integrated transmitters and receivers in radar frequency bands.